\documentclass[prb,aps]{revtex4}
\usepackage{graphicx,color}

\begin{document}

\title{\Large High temperature spin selectivity in a quantum\\
dot qubit using reservoir spin accumulation}

\vspace*{-20mm}
\author{R. Jansen$^{1}$ and S. Yuasa$^1$}
\affiliation{$^1\,$Research Center for Emerging Computing Technologies, National Institute of Advanced Industrial Science and Technology (AIST), Tsukuba, Ibaraki, 305-8568, Japan.}

\begin{abstract}
{\bf
Employing spins in quantum dots for fault-tolerant quantum computing in large-scale qubit arrays with on-chip control electronics requires high-fidelity qubit operation at elevated temperature. This poses a challenge for single spin initialization and readout. Existing schemes rely on Zeeman splitting or Pauli spin blockade with typical energy scales of 0.1 or 1 meV for electron-based qubits, so that sufficient fidelity is obtained only at temperatures around or below 0.1 or 1 K, respectively. Here we describe a method to achieve high temperature spin selectivity in a quantum dot using a reservoir with a spin accumulation, which deterministically sets the spin of a single electron on the dot. Since spin accumulation as large as 10 meV is achievable in silicon, spin selection with electrically adjustable error rates below $10^{-4}$ is possible even in a liquid He bath at 4 K. Via the reservoir spin accumulation, induced and controlled by a nearby ferromagnet, classical information (magnetization direction) is mapped onto a spin qubit. These features provide the prospect of spin qubit operation at elevated temperatures and connect the worlds of quantum computing and spintronics.
}
\end{abstract}

\maketitle

\section{INTRODUCTION}
The development of quantum computers based on spin in quantum dots \cite{review1,review2} has made tremendous advances, from the demonstration of initialization \cite{elzerman,simmons,taruchananomagnet2}, coherent control \cite{petta,koppens,kawakaminanomagnet,taruchananomagnet2} and readout \cite{elzerman,carroll,west,defranceschi} of the single spin quantum state, via two-qubit operations \cite{twoqubit1,twoqubit2}, to architecture design \cite{architecture1,architecture2} and manufacturing with advanced semiconductor technology \cite{CMOSmanufacturing}. For large scale arrays of multi-terminal quantum dot (QD) qubits, the control electronics should be integrated on-chip to avoid an interconnect problem \cite{architecture1,architecture2}. This creates another challenge \cite{yanghot,petithot,warburton}, because balancing the heat generated by the control electronics is difficult at temperatures below 100 mK in a dilution refrigerator, which has limited cooling power. In a liquid He bath at 4.2 K the cooling power is three orders of magnitude higher. Hence, it is indispensable that qubits can operate at higher temperatures \cite{yanghot,petithot,warburton} while at the same time maintaining error rates well below the threshold ($10^{-2}$) for fault-tolerant quantum computing \cite{faulttolerant1,faulttolerant2}.\\
\indent The traditional method \cite{elzerman,simmons,taruchananomagnet2} of initialization/readout of a QD spin qubit, based on Zeeman spin splitting of the dot's energy levels in a magnetic field $B_0$ (Fig. 1a), yields limited fidelity and operation temperature $T$ $<$ 100 mK because the Zeeman splitting is $<$ 0.1 meV for practical values of $B_0$. Therefore, only at very low $T$ selective tunneling of an electron from the lead into the lowest-energy spin state (initialization) occurs. Thermal broadening in the leads at elevated $T$ makes the higher-energy spin state accessible, thus producing initialization errors. A similar issue occurs for readout. Attention has therefore necessarily shifted to the only other available method for initialization/readout, which is based on Pauli spin blockade \cite{pauli1,petta,koppens,pauli2} in double QD qubits. This method, which has been around for more than two decades, has a larger energy scale that is set by the singlet-triplet splitting. For electron-based quantum dots, the singlet-triplet splitting is often comparable to the valley splitting, with reported values of 0.3 - 0.8 meV for Si quantum dots \cite{spinvalley1} and up to about 0.3 meV for the SiGe/Si system \cite{philips}. The readout window is enhanced \cite{philips} for certain multi-electron configurations in which the orbital excited state energy comes into play (an orbital excited state energy of 1 meV was measured for a 4-electron Si quantum dot \cite{orbital}). With Pauli spin blockade, qubit operation at $\sim$1 K was achieved for Si quantum dots with electrons \cite{yanghot,petithot} and holes \cite{warburton} with fidelities that are approaching the fault-tolerance threshold, although some degradation for higher T up to 4 K was also observed \cite{warburton}. Fidelities exceeding the fault-tolerance threshold have recently been reported \cite{noiri,xue,mills}, but only at temperatures below 100 mK. While these are significant advances, one already operates close to the limits of the Pauli spin blockade technique. What is desirable is a method for initialization/readout that can provide a leap forward.

\section{RESULTS AND DISCUSSION}
\noindent {\bf Principle of the approach}\\
\noindent The scheme for the high temperature initialization and readout of QD spins we describe here employs a drain reservoir with a spin accumulation $\Delta\mu = \mu_{D}^{\uparrow} - \mu_{D}^{\downarrow}$. The electrochemical potentials $\mu_{D}^{\uparrow}$ and $\mu_{D}^{\downarrow}$ for the spin-up and spin-down electrons describe the spin-dependent {\it occupation} of the electronic states of the reservoir (Fig. 1a). Because $\Delta\mu$ can be much larger than the thermal broadening, in a wide energy interval between $\mu_{D}^{\uparrow}$ and $\mu_{D}^{\downarrow}$, the reservoir states are completely empty or full, depending on the spin. This provides a thermally robust spin selectivity. The $\Delta\mu$ can be created by electrical \cite{jedema1,tombros,crowellnphys,jansennmatreview,2tmrvanwees,spiesserpra,jansennonlinear,shiraishipra,spiesser90percent}, dynamical \cite{spinpumping1,spinpumping2,spinpumping3} or thermal \cite{slachter,lebreton} spin injection from a ferromagnetic (FM) contact, or via the spin Hall effect \cite{spinhall1,spinhall2,spinhall3}. We consider spin injection by a spin-polarized current $I_{FM}$ across a FM tunnel contact to the reservoir (Fig. 1b). This allows routine creation of spin accumulation, with values as large as 13 meV at 10 K (3.5 meV at 300 K) for n-type Si with Fe/MgO contacts \cite{spiesserpra,jansennonlinear,spiesser90percent}, and 4.1 meV at 300 K in graphene \cite{2tmrvanwees}. The $\Delta\mu$ is proportional to the tunnel spin polarization and to $I_{FM}$ and is thus electrically tunable. The sign of $\Delta\mu$ can be reversed by either reversing the magnetization of the contact, or the current polarity \cite{jansennonlinear} (Fig. 1c). Since $\Delta\mu$ decays with distance from the injection contact on the scale of the spin-diffusion length $L_{SD}$, which is typically a few microns \cite{tombros,crowellnphys,shiraishipra}, the magnetic contact is to be placed near the quantum dot (Fig. 1b). Since a large spin accumulation of holes in p-type semiconductors cannot be achieved due to the much shorter spin-relaxation time for holes \cite{hole1,hole2,hole3,hole4}, the method we describe here is applicable to electron-based qubits.\\
\indent The spin accumulation determines the spin occupation of the QD as follows. The energy $E_d$ of the dot's ground state (twofold degenerate at $B_0$ = 0) is positioned between $\mu_{D}^{\uparrow}$ and $\mu_{D}^{\downarrow}$ with an electrostatic gate (Fig. 1c). If $\Delta\mu > 0$, a spin-down electron can tunnel from the source to the dot, and escape into the drain because $E_d > \mu_{D}^{\downarrow}$. However, as soon as a spin-up electron tunnels onto the dot, it is trapped, because there are no empty spin-up states available in the drain owing to the spin accumulation ($E_d < \mu_{D}^{\uparrow}$), while the empty state the electron left behind in the reservoir is quickly filled up due to fast energy relaxation. If the charging energy of the dot is sufficiently large, Coulomb blockade prevents tunneling of a second electron onto the dot. The current is then blocked and the dot is occupied by a single electron in a $|\uparrow\,>$ quantum state (Fig. 1c). For negative spin accumulation ($\mu_{D}^{\uparrow} < \mu_{D}^{\downarrow}$), the situation is similar, except that now the single electron has a $|\downarrow\,>$ quantum state. The spin accumulation thus determines the spin state of a single electron on the dot. Exchange coupling between the dot and the reservoir is not required.

\begin{figure}[htb]
\centering
\hspace*{0mm}\includegraphics*[width=138mm]{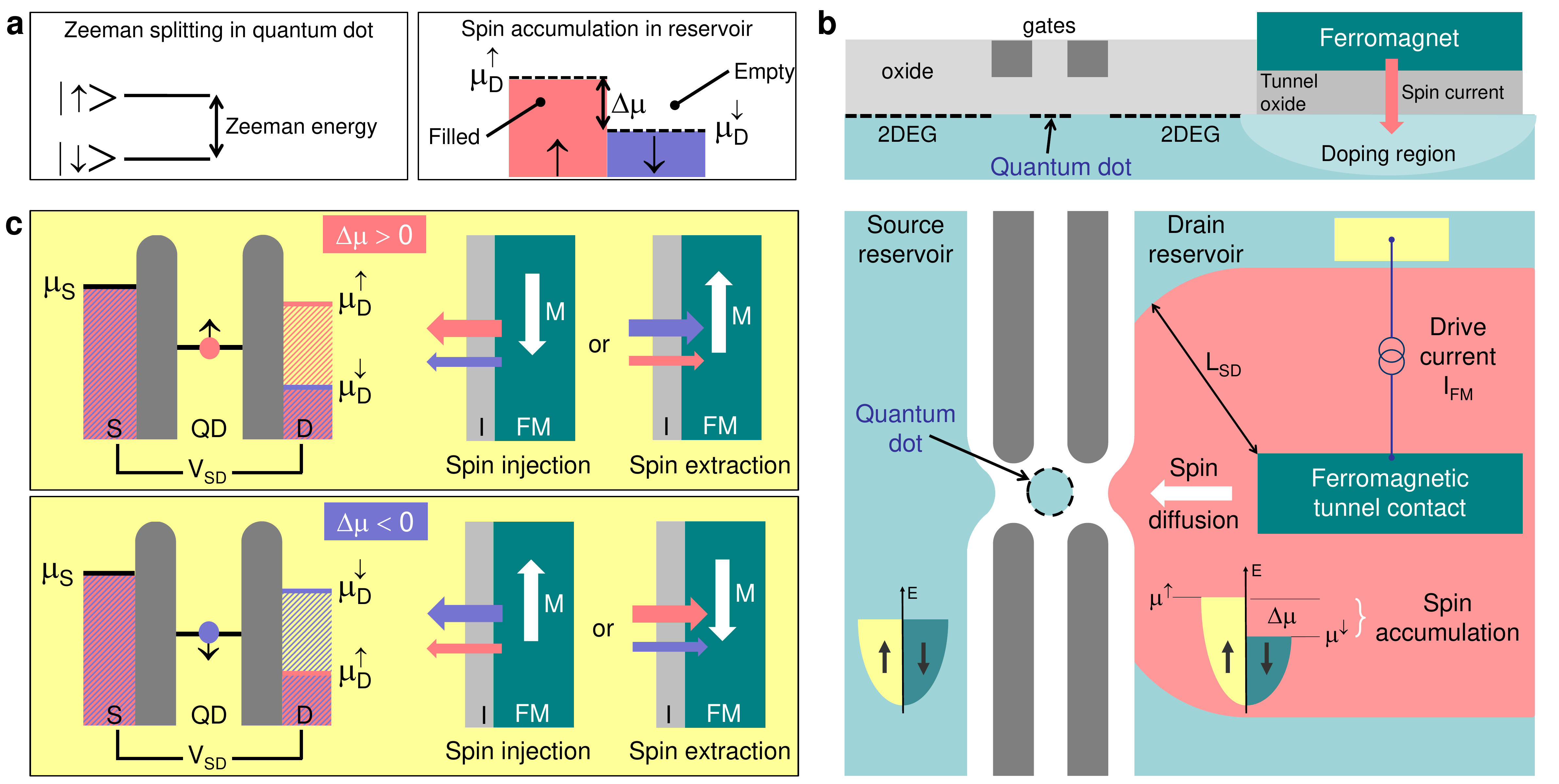}
\caption{{\bf Quantum dot spin qubit with reservoir spin accumulation.} {\bf a}, Comparison between the Zeeman spin splitting of the quantum dot states (the traditional method to achieve spin selectivity in a quantum dot) and the method described here, which employs a reservoir with a spin accumulation $\Delta\mu$, denoting the spin-dependent occupation of the electronic states. {\bf b}, Basic device layout showing the quantum dot defined by electrostatic gates (grey strips). Tunnel barriers (white areas) separate it from the source and drain reservoirs, which consist of a two-dimensional electron gas (2DEG). In the drain reservoir a spin accumulation $\Delta\mu$ is created by an electrical spin current across a ferromagnetic tunnel contact (drive current $I_{FM}$). The induced spin accumulation spreads out (pink area) in the reservoir away from the magnetic contact by spin diffusion (length scale: spin-diffusion length $L_{SD}$). {\bf c}, Energy diagrams showing the spin orientation of a single electron on the quantum dot, as it is determined by the spin accumulation. Positive (negative) $\Delta\mu$ yields a single electron trapped on the quantum dot in a spin-up $|\uparrow\,>$ (spin-down $|\downarrow\,>$) quantum state. The sign of $\Delta\mu$ is controlled by the direction of the magnetization (white arrows) of the ferromagnet (green, label FM) and/or by the polarity of the drive current across the tunnel barrier (light grey, label I), which yields spin injection into or spin extraction from the drain reservoir. The spin-polarized flow from the FM tunnel contact to the drain reservoir is indicated by pink (spin up) and violet (spin down) arrows.}
\label{fig1}
\end{figure}

\noindent {\bf Model description}\\
\noindent In order to confirm the described behavior and quantify the characteristics, we compute the current $I_{SD}$ through the quantum dot and the spin-dependent occupation probabilities $N^{\uparrow}$ and $N^{\downarrow}$ of the dot using rate equations \cite{barnas,eques} (Supplementary Note 1). Single-electron tunneling events induce transitions between the QD states (empty dot / single spin-up electron / single spin-down electron / doubly-occupied singlet state). The transition energy $\Delta_S$ from the ground to the singlet state includes the Coulomb charging energy. The various tunneling rates are functions of the spin-{\em in}dependent tunnel coupling $\Gamma^S$ and $\Gamma^D$ of the dot to the source and drain reservoir, respectively, and of the Fermi-Dirac distribution factors. The latter depend on $E_d$ relative to the source ($\mu_S$) and the spin-dependent drain electrochemical potentials ($\mu_{D}^{\uparrow}$, $\mu_{D}^{\downarrow}$). The source reservoir is taken to be at ground potential. The source-drain voltage $V_{SD}$ shifts the drain electrochemical potentials by $-eV_{SD}$ and shifts the quantum dot level according to $E_d = \epsilon_d-xeV_{SD}$,
with $\epsilon_d$ the dot level energy at $V_{SD}$ = 0, and $e$ the electron charge. The fraction $x$ is determined by the relative values of $\Gamma^S$ and $\Gamma^D$ (we have taken $x$ = 1/2). Spin flips in the quantum dot are included with a rate $\Gamma_1 = \hbar/(2eT_1)$, as governed by the longitudinal spin-relaxation time $T_1$, which is set to 1 ms ($\hbar$ is the reduced Planck's constant). In some cases, a Zeeman splitting of the quantum dot level is included. The formalism considers quantum dot spin states that are collinear with the magnetization of the ferromagnet, which sets the quantization axis of the spin accumulation. This is sufficient for the calculation of the spin selectivity and the error rates. Transverse spin components are needed to describe the non-collinear spin states that appear during qubit spin manipulation by spin resonance. Also, orbital excited states are not included (as explained in Supplementary Note 1).\\
\\
\noindent {\bf Spin selection with reservoir spin accumulation}\\
\indent The calculated $I_{SD}$ exhibits a typical Coulomb staircase, which is modified when $\Delta\mu$ is non-zero (Fig. 2a). More importantly, in the presence of a spin accumulation, a spin-dependent dot occupation is produced in a wide $V_{SD}$ interval, where $N^{\uparrow}$ and $N^{\downarrow}$ are either 1 or 0. A more complete picture is obtained from the stability maps produced by varying $V_{SD}$ and $\epsilon_d$ (Fig. 2b). Without spin accumulation, the current is zero for large positive $\epsilon_d$, where the dot energy level is not accessible, and in the central so-called Coulomb diamond (Fig. 2b, dotted white lines) due to Coulomb blockade. However, because $\Delta\mu$ = 0 and $B_0$ = 0, we have $N^{\uparrow}$ = $N^{\downarrow}$ everywhere. In the presence of a spin accumulation, the boundaries of the central Coulomb diamond are changed, and two regions labeled $\alpha$ and $\beta$ appear in which the dot occupation is highly spin dependent (Fig. 2b). For $\Delta\mu$ = $+$2 meV, we obtain two regions with $N^{\uparrow}$ = 1 and $N^{\downarrow}$ = 0 while also $I_{SD}$ = 0, implying that a single electron in a $|\uparrow\,>$ quantum state is trapped continuously on the dot. When $\Delta\mu$ is reversed to $-$2 meV, a single electron with the opposite spin state ($|\downarrow\,>$) is trapped on the dot ($N^{\uparrow}$ = 0 and $N^{\downarrow}$ = 1). The calculations establish that the reservoir spin accumulation determines the spin quantum state of a single electron on the dot, which provides a scheme for spin qubit initialization and readout. Noteworthy is that $\Delta\mu$ is controlled by the magnetization of the FM contact. This means that a classical bit of information, represented by the magnetization direction, is mapped onto the quantum state of a single electron spin.

\begin{figure}[htb]
\centering
\hspace*{0mm}\includegraphics*[width=158mm]{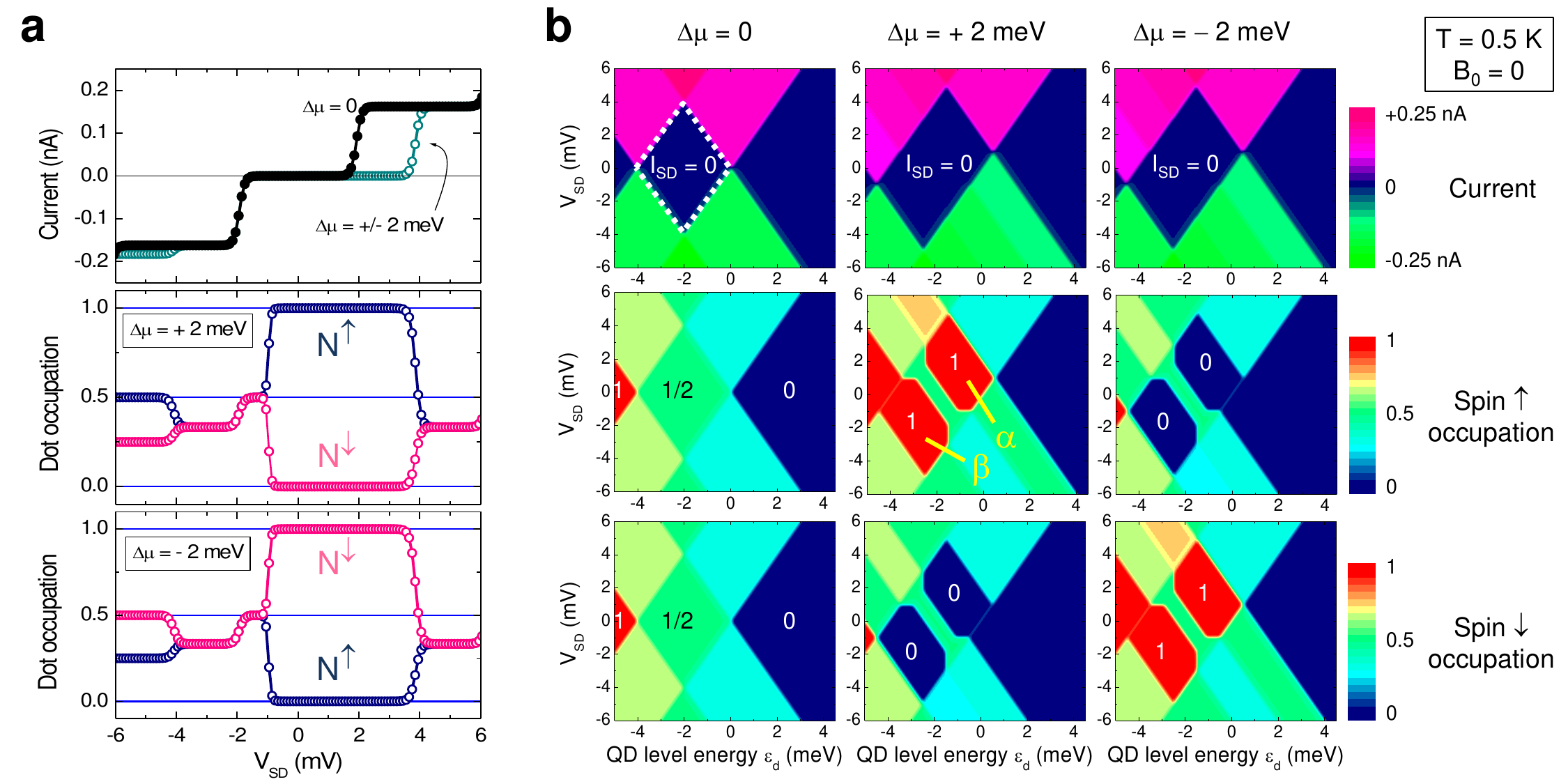}
\caption{{\bf Current and spin occupation of a quantum dot with reservoir spin accumulation.} {\bf a}, Calculated current and dot occupations $N^{\uparrow}$ and $N^{\downarrow}$ as a function of the source-drain voltage $V_{SD}$ for $\Delta\mu$ = 0, $+$2 meV and $-$2 meV. The energy $\epsilon_d$ of the singly-occupied quantum dot ground state at $V_{SD}$ = 0 was taken to be $-$1 meV. {\bf b}, Stability maps of the current, $N^{\uparrow}$ and $N^{\downarrow}$ as a function of $\epsilon_d$ and $V_{SD}$. With spin accumulation, there are two regions (labeled $\alpha$ and $\beta$) in which $N^{\uparrow}$ = 1 and $N^{\downarrow}$ = 0 (for $\Delta\mu >$ 0) or $N^{\uparrow}$ = 0 and $N^{\downarrow}$ = 1 (for $\Delta\mu <$ 0), while $I_{SD}$ is zero. In these regions, a single electron is trapped on the quantum dot in a spin-up $|\uparrow\,>$ or a spin-down $|\downarrow\,>$ quantum state, depending on the sign of $\Delta\mu$. Calculations performed at $T$ = 0.5 K and without Zeeman splitting ($B_0$ = 0) using a ground state to singlet state transition energy $\Delta_S$ of 4 meV. Transitions to the triplet states are not included here. On the horizontal axes we have used $\epsilon_d$ (the dot level energy at $V_{SD}$ = 0). Using the actual dot level energy $\epsilon_d - xeV_{SD}$ would skew the stability maps. }
\label{fig2}
\end{figure}

\indent While triplet states play no role for region $\alpha$, they are required to properly describe region $\beta$, as explained below. We therefore included triplet states in the calculations (Supplementary Note 1). The stability maps for different values of $\Delta\mu$ (Fig. 3a) show that the boundaries of region $\alpha$ depend on $\Delta\mu$. The upper right boundary of the central Coulomb diamond, which is defined by the dot level energy $E_d$ relative to the drain electrochemical potential, splits into two boundaries (solid pink lines in Fig. 3b) because $\mu_{D}^{\uparrow} \neq \mu_{D}^{\downarrow}$. Spin selectivity then appears when the ground state lies between $\mu_{D}^{\uparrow}$ and $\mu_{D}^{\downarrow}$, as already discussed above. For region $\beta$, the origin of the spin selectivity is different (Fig. 3c). For region $\beta$, the ground state lies below the reservoir electrochemical potentials, while a transition to the doubly-occupied singlet state (label S) requires an energy between $\mu_{D}^{\uparrow}$ and $\mu_{D}^{\downarrow}$. This also produces a single electron in a $|\uparrow\,>$ quantum state (for $\Delta\mu > 0$). A spin-down electron is not trapped in the ground state because a spin-up electron can tunnel in from the drain. From the resulting singlet state, an electron of either spin can tunnel out. If a spin-up electron tunnels out, we return to the starting configuration, and the process repeats itself until the spin-down electron tunnels out from the singlet state. This leaves behind an electron in a $|\uparrow\,>$ ground state. This is trapped if (i) the singlet state is energetically not accessible for a spin-down electron, and (ii) a spin-up electron cannot tunnel into the dot to create a triplet state (label T), which requires an additional energy given by the singlet-triplet transition energy $\Delta_{ST}$. If $\Delta\mu > \Delta_{ST}$, as we consider here, the lower boundary of region $\beta$ occurs when $\mu_{D}^{\uparrow}$ crosses the triplet energy and a non-zero current appears. The spin selectivity for region $\beta$ then results from the combination of the spin accumulation and Pauli spin blockade. Therefore, the relevant energy scale is $\Delta_{ST}$, whereas it is $\Delta\mu$ for region $\alpha$.

\begin{figure}[htb]
\centering
\includegraphics*[width=169mm]{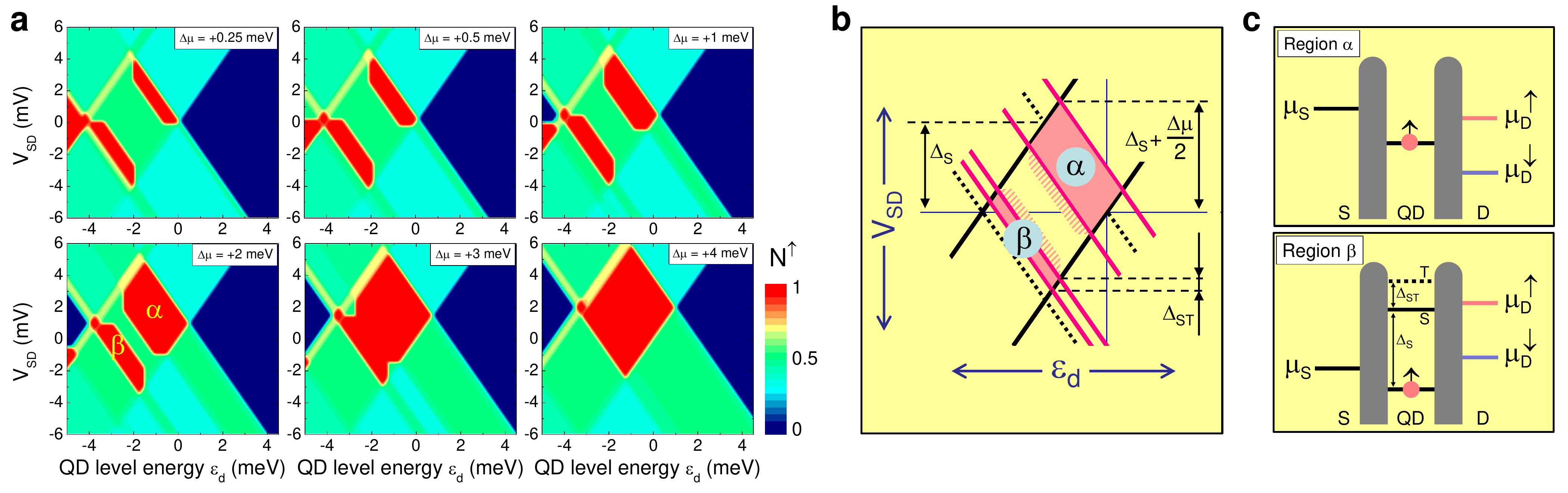}
\caption{{\bf Operation window for spin selection in a quantum dot with reservoir spin accumulation.} {\bf a}, Stability maps of $N^{\uparrow}$ as a function of $\epsilon_d$ and $V_{SD}$, showing the evolution of the high spin-selectivity regions $\alpha$ and $\beta$ as the spin accumulation is increased from $+$0.25 meV to $+$4 meV (calculated for $T$ = 0.5 K, $B_0$ = 0, $\Delta_S$ = 4 meV and $\Delta_{ST}$ = 0.5 meV). {\bf b}, Sketch of the regions $\alpha$ and $\beta$ in the presence of a spin accumulation. The two boundaries marked by the solid black lines do not change, whereas the two boundaries marked by the dotted black lines are modified (pink lines) for a non-zero $\Delta\mu$. The actual boundaries of $\alpha$ and $\beta$ extend beyond the pink lines, as indicated by the dashed pink areas (Supplementary Note 2). {\bf c}, Energy diagrams showing the spin orientation of a single electron on the quantum dot for regions $\alpha$ and $\beta$. Indicated are the energies $\Delta_S$ and $\Delta_{ST}$ for transitions from, respectively, the ground to the singlet state (label S) and from the singlet to the triplet state (label T). In region $\alpha$, the quantum dot spin is determined by the spin accumulation. For region $\beta$ the mechanism is different, and the quantum dot spin is determined by a combination of the spin accumulation and Pauli spin blockade.}
\label{fig3}
\end{figure}

\noindent {\bf High temperature spin selection}\\
\indent The $\Delta\mu$ provides a large energy scale that makes the spin selectivity robust against temperature. Region $\alpha$ with $N^{\uparrow} \approx$ 1 and $N^{\downarrow} \approx$ 0 persists even at 6 K for a $\Delta\mu$ of $+$3 meV (Fig. 4a). These results were obtained with a 70 $\mu$eV Zeeman splitting that favours the spin-down state, opposite to the spin accumulation. But because $\Delta\mu >>$ 70 $\mu$eV, the spin accumulation defines the quantum dot spin. For region $\beta$, the $N^{\uparrow}$ decays more rapidly with increasing $T$ because the relevant energy scale is $\Delta_{ST}$, which was set to 0.5 meV. The decay of $N^{\uparrow}$ is accompanied by a finite current (of spin-up electrons via the triplet states), making region $\beta$ unsuitable for qubit operation at 4 K. The boundary of region $\beta$ for $N^{\downarrow}$ is determined by the relative position of $\mu_{D}^{\downarrow}$ and the singlet state, and thus by $\Delta\mu$, so that $N^{\downarrow}$ for region $\beta$ remains $\approx$ 0 even at 6 K. For a more quantitative analysis, we extract the error rates from region $\alpha$. The error rate is defined here as the probability for finding the spin state opposite to what was desired, i.e., the error rate is $N^{\downarrow}/N^{\uparrow}$ for $\Delta\mu >$ 0 and $N^{\uparrow}/N^{\downarrow}$ for $\Delta\mu <$ 0. Without spin accumulation, the error rate rises quickly and surpasses the fault-tolerance threshold for $T >$ 100 mK (Fig. 4b). With a spin accumulation, the error rate remains much smaller at higher $T$. At 4.2 K, a $\Delta\mu$ of about 2 or 3 meV yields an error rate of 10$^{-2}$ or 10$^{-3}$, which is below the fault-tolerance threshold (red line). For larger spin accumulation, higher $T$ and/or smaller error rates are possible. The latter reduces the qubit array size required for computation \cite{faulttolerant1,faulttolerant2}.

\begin{figure}[htb]
\centering
\includegraphics*[width=175mm]{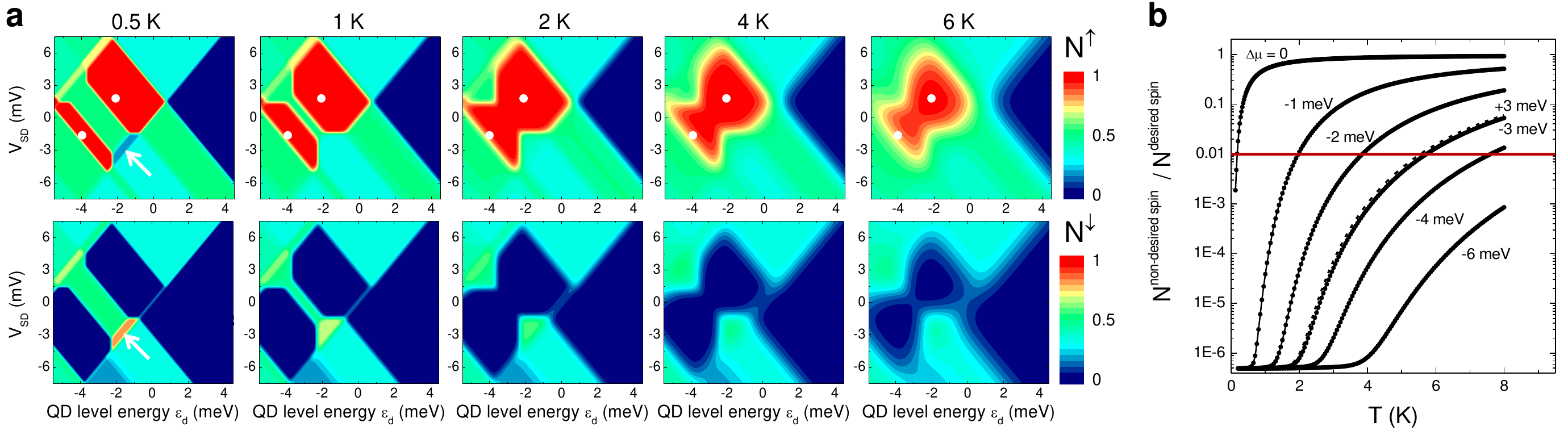}
\caption{{\bf Robustness of spin selectivity at higher temperature.} {\bf a}, Quantum dot spin occupations $N^{\uparrow}$ and $N^{\downarrow}$ for various temperatures (0.5, 1, 2, 4 and 6 K), showing that with a reservoir spin accumulation of $+$3 meV, the high spin selectivity is maintained above 4 K despite the thermal broadening. Calculations for $\Delta_S$ = 6 meV and $B_0$ = 0.6 T, producing a Zeeman splitting that opposes the spin selected by the spin accumulation. The Zeeman splitting controls the spin occupation in a narrow region near the lower right edge of the Coulomb diamond (white arrows) at $T$ = 0.5 K, but this disappears quickly at higher $T$ (see also the stability maps vs. $T$ for $\Delta\mu$ = 0 in the Supplementary Note 3). The white dots in the top panels are guides to the eye and are in the exact same location for each panel. Note that the lower boundary of region $\beta$ is different for $N^{\uparrow}$ and $N^{\downarrow}$ (see Supplementary Note 3). {\bf b}, Temperature variation of the error rate, represented as the ratio of the occupation probabilities for the non-desired and the desired spin, as extracted from a location near the white dot in region $\alpha$ (at least 0.5 mV away from the boundary). When $\Delta\mu$ = 0, only the small Zeeman splitting for a typical 0.6 T magnetic field is active and the error rates quickly rise above the threshold (red line) for fault-tolerant quantum computing, thus restricting the operation temperature to $<$ 100 mK. In contrast, with a reservoir spin accumulation of 2-3 meV, the error rates remain below the threshold even at the liquid He bath temperature of 4.2 K. Even higher temperatures and lower error rates are possible for larger spin accumulation.}
\label{fig4}
\end{figure}

\noindent {\bf Implementation in spin qubits}\\
\noindent We describe a basic sequence for the initialization, control and readout of a quantum dot qubit with a reservoir spin accumulation (Fig. 5). In the simplest scheme, the magnitude and sign of the spin accumulation are kept constant throughout the entire sequence. After emptying the dot, $V_{SD}$ and $\epsilon_d$ are adjusted so as to reach region $\alpha$ in the stability diagram and thereby initialize a single electron with a well-defined spin on the dot. The selected spin state is determined by the sign of $\Delta\mu$. A large operation window (region $\alpha$) exists for $V_{SD}$ and $\epsilon_d$, making the system insensitive to errors produced by potential variations. In the next step, the quantum dot level is moved down so as to obtain a deep Coulomb blockade for spin manipulation by established spin resonance methods \cite{koppens,kawakaminanomagnet,taruchananomagnet2}. Care has to be taken that during this stage the qubit spin is not disturbed by the spin accumulation in the drain reservoir (see the Supplementary Discussion). In the final step, $V_{SD}$ and $\epsilon_d$ are adjusted to their readout values. The spin accumulation enables readout because only one type of electron can tunnel out into the drain (spin down for the case of $\Delta\mu > 0$ depicted in Fig. 5). The tunneling event can be detected via established charge sensing techniques \cite{elzerman,carroll,west,defranceschi}.

\begin{figure}[htb]
\centering
\includegraphics*[width=68mm]{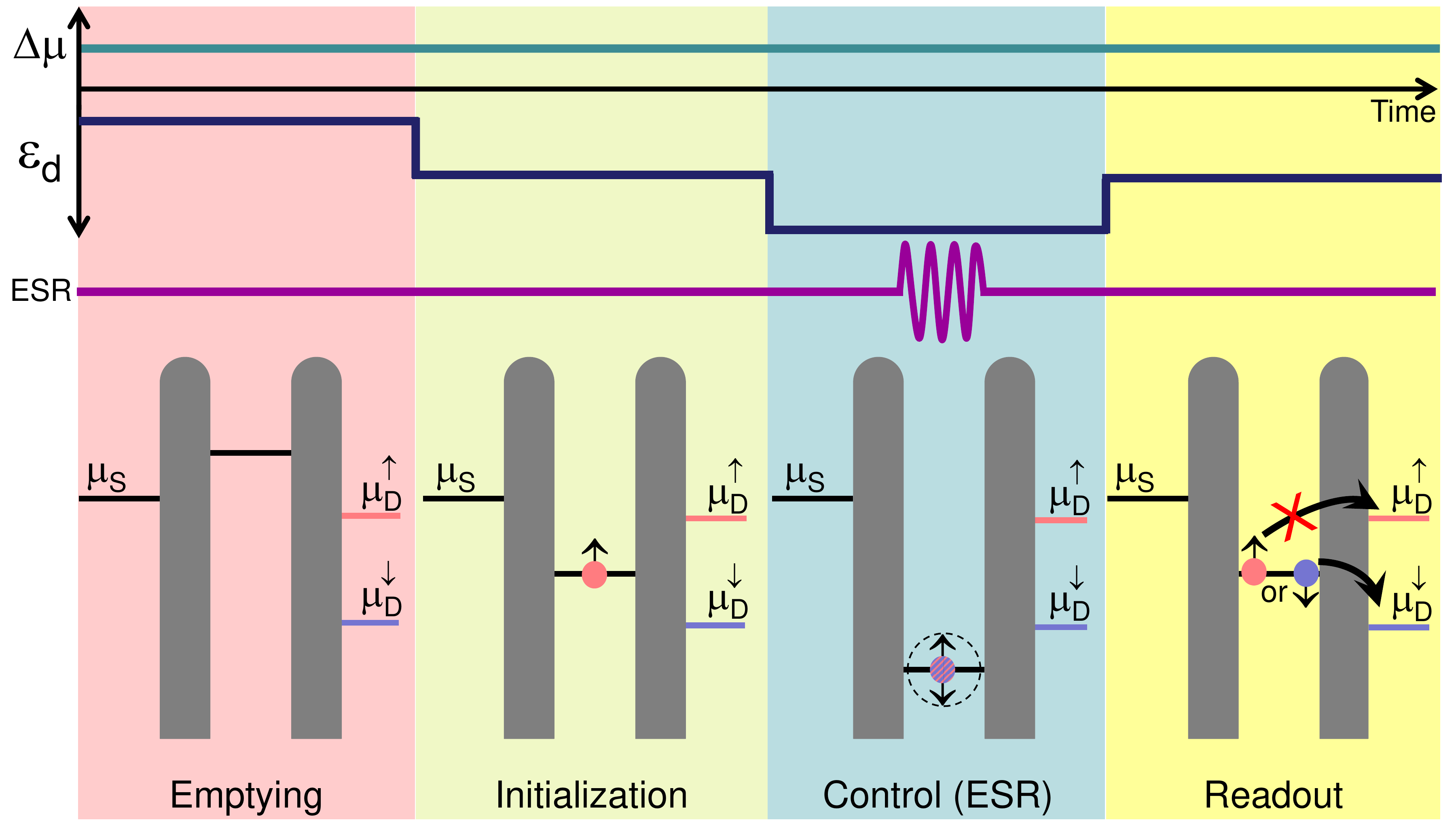}
\caption{{\bf Basic operation sequence of a quantum dot spin qubit with reservoir spin accumulation.} In the simplest scheme, the magnitude and sign of the spin accumulation are kept constant throughout the entire sequence. The initial spin state of the quantum dot is set by the reservoir spin accumulation and by choosing appropriate settings for $\epsilon_d$ and $V_{SD}$. For the next step, spin control by established electron spin resonance (ESR) techniques for spin qubits, $\epsilon_d$ is adjusted so as to obtained a deep Coulomb blockade. For readout, the spin accumulation is used to produce spin selectivity. If after the spin control by spin resonance a $|\uparrow\,>$ quantum state is obtained, the electron cannot tunnel into the drain reservoir owing to the non-zero $\Delta\mu$. If a $|\downarrow\,>$ quantum state is obtained after spin resonance, the electron can tunnel into the drain, and the associated charge movement can be detected by established charge sensing techniques.}
\label{fig5}
\end{figure}

\indent The large spin accumulation offers high spin selectivity for initialization and readout. For the overall readout fidelity, other factors also come into play, including errors that might occur in the charge sensing of the readout transition. Such errors depend on the charge sensing method and on other design details, which are beyond the scope of this work. Since $\Delta\mu$ is adjustable electrically (via the current through the magnetic contact to the drain), the spin selectivity can be tuned to levels for which the overall fidelity is limited only by the spin control process. Variations of the operation sequence presented in Fig. 5 are possible. For instance, the spin that is selected for initialization and readout can be changed by reversing the magnetization of the ferromagnetic contact to the drain. However, local magnetization control for each reservoir contact requires additional elements and increases the complexity and footprint of the qubit design. One might also consider active control of $\Delta\mu$, for instance switching it off during the spin resonance manipulation. This needs to be done in a way that does not disturb the spin qubit (see the Supplementary Discussion for more information). It is emphasized that reversal of the spin accumulation does not constitute coherent spin control because the originally trapped electron tunnels out and is replaced by a different electron with opposite spin. It has been proposed that coherent manipulation of a qubit spin by scattering off "flying" spins in the reservoir is possible if an exchange coupling exists between the qubit and the reservoir spins \cite{datta}.\\
\indent Surely, the use of a drain reservoir with a ferromagnet to create the spin accumulation introduces additional complexity, increases the size of the spin qubit, affects the scalability, and may also modify the noise during qubit operation. At the same time, the described method enables high spin selectivity, enhanced robustness at higher temperature, with a wide operation window for $V_{SD}$ and $\epsilon_d$. Also, the approach is conceptually simple as it uses one quantum dot with a single electron, instead of the multi-electron double-dot devices used for Pauli spin blockade \cite{yanghot,petithot,warburton,pauli1,petta,koppens,pauli2,philips}. In future work, it is of importance to investigate the required modifications to the design of the qubit array and assess whether the envisioned benefits of the described method outweigh the potential drawbacks. Some initial considerations are given in Supplementary Note 4 and the Supplementary Discussion, which also describe aspects related to the practical implementation. This includes the feasibility to induce a sufficiently large spin accumulation in the 2DEG reservoir, the applicability to different qubit material systems (not only quantum dots in Si/SiO$_2$ 2DEG's but also SiGe/Si with a buried 2DEG, or dopant-based spin qubits), the expected dimensions of a reservoir with a ferromagnetic contact, and the drive current $I_{FM}$ ($\sim$ 3 nA) and power ($\sim$ 20 pW) required to achieve the desired $\Delta\mu$. We note that nanomagnets are already widely used \cite{yanghot,taruchananomagnet2,kawakaminanomagnet,philips,taruchananomagnet1} in spin qubits to produce an inhomogeneous magnetic field for electrically driven spin resonance (EDSR). It is of interest to examine whether the ferromagnetic tunnel contact to the drain reservoir can have a dual function (create spin accumulation and assist in producing the magnetic field gradients required for EDSR).\\
\indent In summary, we have described a scheme for high temperature initialization and readout of a spin qubit. It is based on a reservoir with a spin accumulation, which controls the spin state of a single electron on a quantum dot. The spin accumulation is induced electrically by spin injection from a nearby ferromagnetic tunnel contact. The method enables operation at elevated $T$, even above 4.2 K, with error rates below $10^{-4}$. The error rates are electrically adjustable and the operation window for $V_{SD}$ and $\epsilon_d$ is wide, making the approach insensitive to potential variations. Via the spin accumulation, a classical bit of information (magnetization direction) is mapped onto the quantum spin state of a single electron. This connects spin qubits and quantum computing to spintronics.

\section{Data availability}
The authors declare that the data supporting the findings of this study are available within the paper and its Supplementary Information files.

\section{Code availability}
All the codes used in this work are available upon request to the authors.

\section{Acknowledgement}
The authors thank A. Spiesser for initiating investigations into the use of spin current for spin-based quantum computing, which stimulated the authors to explore other ways to use spintronic phenomena in quantum dot qubits. We also thank S. Kawabata for useful discussion.

\section{Author contributions}
RJ designed the concept, performed the calculations and analysis of the results, and wrote the draft and the final manuscript. SY reviewed and gave input on the manuscript and provided overall supervision.

\section{Competing interests}
The authors declare no competing interests.

\end{document}